# Three-dimensional Self-assembled Columnar Arrays of AlInP Quantum Wires for Polarized Micron-sized Amber Light Emitting Diodes

Andrea Pescaglini,[a] Agnieszka Gocalinska,[a,*] Silviu Bogusevschi,[a,b] Stefano T. Moroni,[a] Gediminas Juska,[a] Enrica E. Mura,[a] John Justice,[a] Brian Corbett,[a] Eoin O'Reilly,[a,b] Emanuele Pelucchi[a]

[a] Tyndall National Institute – University College Cork, Lee Maltings, Cork, Ireland

[b] Department of Physics, University College Cork, Ireland

*A three-dimensional ordered and self-organized semiconductor system emitting highly-polarized light in the yellow-orange visible range (580-650 nm) is presented, comprising self-assembled in-plane AlInP wires vertically stacked in regularly-spaced columns. More than 200 wires per column without detectable defect formation could be stacked. Theoretical simulations and temperature-dependent photoluminescence provided a benchmark to engineer multilayered structures showing internal quantum efficiency at room temperature larger than comparable quantum wells emitting at similar wavelengths. Finally, proof-of-concept light emitting diodes (LED) showed a high degree of light polarization and lower surface parasitic currents than comparable quantum well LEDs, providing an interesting perspective for high-efficiency polarized yellow-orange light emitting devices.*

Light emitting diodes (LEDs) are currently driving innovation in many technological areas, such as displays for TV/monitors, mobiles and laptops, lighting and automotive, optical interconnections, logic and sensing.[1-4] A key factor for success is the ability to produce light with high efficiency and with a very small footprint, i.e. fundamental properties compatible with the demanding requirements of compactness and integrability. As a consequence, considerable effort has been dedicated to the investigation of novel materials and designs for LED devices, aiming to improve efficiency, spectral properties and to add integrated functionalities. [5-14]

However, some fundamental issues in light emitting diode technology are still outstanding. For instance, semiconductor-based III-V LEDs show a well-known drop in efficiency at wavelengths in the range of 550-640 nm (yellow-red optical band of the spectrum) in both nitride-based and phosphorus-based compounds.[15] In III-P compounds this is linked to poor electron confinement (related to the direct-to-indirect bandgap transition for Al:Ga ratio above ~53:47% in AlGaInP lattice-matched to GaAs,[15] which result in a severe reduction in the radiative recombination efficiency), while in InGaN materials the increasing In content, required to shift the emission from blue towards the yellow, introduces a larger number of structural defects, segregation and strong quantum mechanical Stark effects that seem unavoidable detrimental factors for optical properties.[16-19] Moreover, traditional quantum well LEDs can show a strong reduction in current-to-light conversion efficiency due to non-radiative recombination paths created by surface states (particularly relevant in LEDs in the micron size range), a very relevant limiting factor for truly microLED technology implementation. It should also be added that polarized light emission, critically required in a variety of applications such as LCD screens and environmental lighting,[20-22] is presently achieved only by additional structures such as gratings, filters and plasmonic nanostructures that increase the complexity of the device fabrication and reduce the external quantum efficiency.[23-30]

A viable approach able to mitigate these outstanding issues could be the exploitation of carrier confinement in low dimensional nanostructures. For instance, epitaxial growth has been used to self-assemble high density and high crystal quality quantum dots (QDs) providing high temperature stability and low threshold currents in lasers and LEDs.[31-33] Nevertheless this is a paradigm that critically requires no (defect-related) lattice relaxation. AlGaInP alloys for red emitters do not offer issues in terms of lattice relaxation, as lattice-matching to GaAs can be constantly maintained while varying the Al content to shift the emission from red (650 nm for simple InGaP lattice-matched to GaAs) to yellow, maintaining high crystal quality without any polar crystallographic direction. We observe, on the other hand, that although the quasi 0-dimensionality of QDs allows maximizing the carrier quantum confinement, polarized light for surface emission cannot be directly produced due to a generally high symmetry of the crystallographic structures, as it is also the case in the traditionally exploited quantum wells.

In this regard, the one-dimensionality of wire-like nanostructures would allow the unique coexistence of better carrier confinement and spontaneous polarized emission.[34] However, reliable and scalable methods to grow embedded self-assembled wires without pre-patterned templates (such as V-grooves),[35, 36] and with a design optimized for yellow-orange light emission are, to date, missing.

Here we present a reproducible growth process to realize ordered self-assembled multilayered in-plane AlInP/AlGaInP quantum wires (SMWRs) grown on a GaAs substrate. The wires, with lateral size of ~10-25 nm, center-to-center spacing in the growth plane of ~50 nm and a few microns in length, can be organized in vertical columns of up to 200 wires per column, and show light emission in the yellow-orange visible range (580-650 nm). Temperature-dependent photoluminescence spectroscopy alongside a 8-band **k·p** model simulations were exploited to unveil the fundamental physical processes in light production, including the influence of escape processes and band alignments, and to estimate the internal quantum efficiency (IQE).

We also show that the IQE of a 100 layer SMWR structure can be critically improved by adding a quantum well (QW) coupled to the SMWR system (QW-SMWR), functioning as a carrier injector. When compared to a (comparable, but by no means optimized) AlGaInP five QW structure, the IQE of the QW-SMWR hybrid system outperformed the QW structure in the entire range of temperature explored (4-300 K), exhibiting up to one order of magnitude higher internal quantum efficiency (at 100 K). Finally, prototype LED devices were characterized showing low turn-on voltage, lower surface recombination currents compared to QW LEDs and highly polarized in-plane light emission. The results obtained from the QW-SMWR system show great potential in terms of design flexibility and emission efficiency, opening up interesting perspectives towards development of cryogenic and room temperature optoelectronic devices.

All epitaxial samples discussed were grown in a high purity MOVPE[37-39] commercial horizontal reactor (AIX 200) at low pressure (80 mbar) with purified $N_2$ as carrier gas. The precursors were trimethylindium (TMIn), trimethylgallium (TMGa), trimethylaluminium (TMAl), diethylzinc (DEZn) arsine ($AsH_3$), phosphine ($PH_3$) and disilane ($Si_2H_6$). The samples' designs consisted in 100 nm $Al_{0.75}Ga_{0.25}As$, 100 nm $(Al_{0.8}Ga_{0.2})_{0.52}In_{0.48}P$ followed by alternating layers of $(Al_{0.6}Ga_{0.4})_{0.52}In_{0.48}P$ barrier and of $Al_xIn_{1-x}P$ (0.15<x<0.30) with nominal thickness of 3 nm and 0.4 nm respectively. Variation to this design will be described in the text. All growth temperatures (T) quoted have been estimated by emissivity corrected pyrometry. All samples had a homoepitaxial GaAs 100 nm thick buffer grown prior to the described structure and were capped with 200 nm of $(Al_{0.8}Ga_{0.2})_{0.52}In_{0.48}P$. Growth conditions for the AlGaInP layers were: V/III ratio ~630, growth rate ~0.4 nm/s; growth conditions for the AlInP layers were: V/III ratio 290, growth rate 0.3 nm/s; samples were grown at growth T of 720 °C on semi-insulating (100) 6° ± 0.02° off towards [111] A (unless stated otherwise). The LED devices were grown on a n-doped GaAs substrate 6° ± 0.02° off towards (111) A. The structure consisted in 100 nm $Al_{0.75}Ga_{0.25}As$:Si ($n\sim10^{18}cm^{-3}$), 100 nm $(Al_{0.8}Ga_{0.2})_{0.52}In_{0.48}P$:Si ($n\sim10^{18}cm^{-3}$) followed by alternating layers of $Al_xIn_{1-x}P$ and of $(Al_{0.6}Ga_{0.4})_{0.52}In_{0.48}P$, as previously described. All samples had a homoepitaxial GaAs:Si ($n\sim10^{18}cm^{-3}$) 100 nm thick buffer grown prior to the described structure and were finalized with 100 nm of $(Al_{0.8}Ga_{0.2})_{0.52}In_{0.48}P$, 100 nm of $(Al_{0.8}Ga_{0.2})_{0.52}In_{0.48}P$:Zn ($n\sim10^{18}cm^{-3}$), 100 nm AlInP:Zn ($n\sim10^{18}cm^{-3}$), 10 nm InGaP:Zn ($n\sim10^{18}cm^{-3}$) and 10 nm of GaAs:Zn ($n\sim10^{18}cm^{-3}$). Standard optical lithography, dry-etch and deposition of Ti/Pt/Au (20/30/200 nm) metal contact on the p-side were used to fabricate the LED devices (further details in the Supporting Information (SI)).

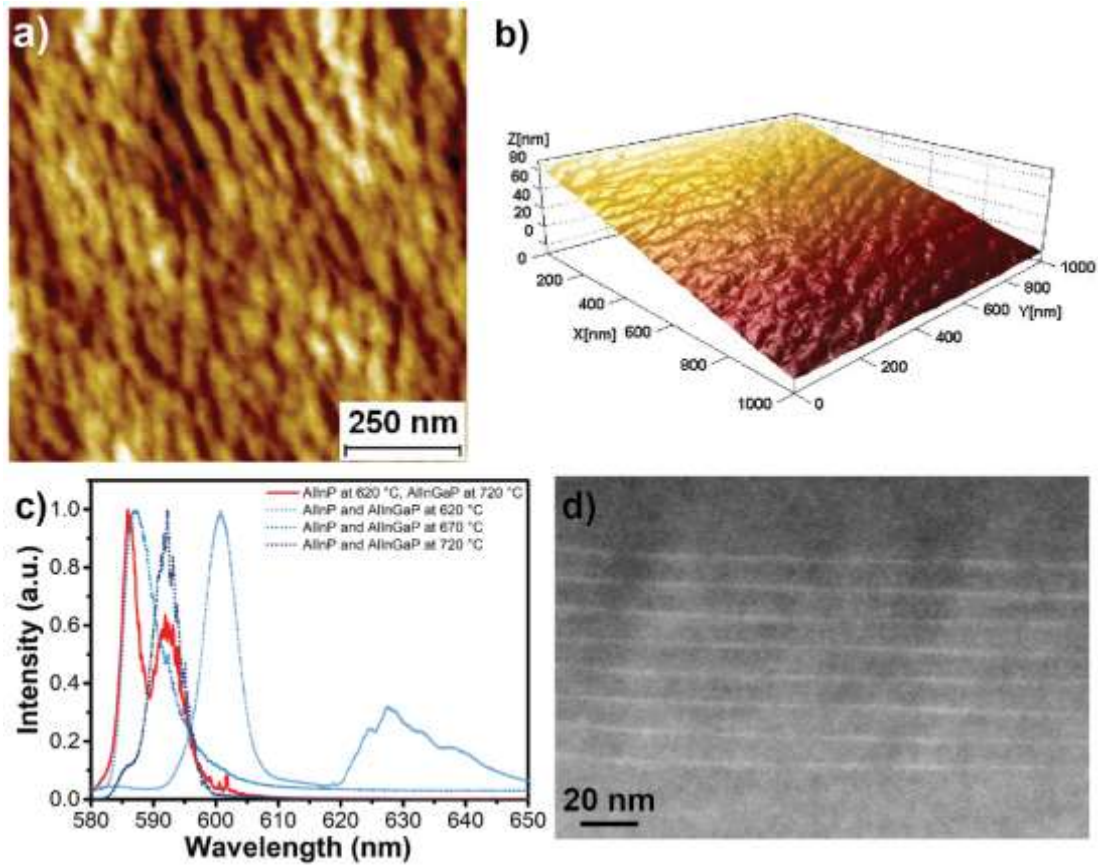

**Figure 1**: AFM images (a) signal amplitude and (b) reconstructed 3D height image of a single un-capped layer of 0.4 nm (nominal) of $Al_{0.2}In_{0.8}P$ grown on a 6° off GaAs substrate. (c) photoluminescence spectra (at 8K) of single capped $Al_{0.2}In_{0.8}P/(Al_{0.6}Ga_{0.4})_{0.52}In_{0.48}P$ layer grown at 620 (light blue dots), 670 (blue dots), 720 (dark blue dots) °C, respectively, compared to a sample in which $(Al_{0.6}Ga_{0.4})_{0.52}In_{0.48}P$ layers were grown at 720 °C and AlInP layers at 620 °C (red line). (d) Cross-section TEM of a SMWR with 8 $Al_{0.2}In_{0.8}P$ layers separated by 10 nm $(Al_{0.6}Ga_{0.4})_{0.52}In_{0.48}P$ barriers.

Deposition of lattice-mismatched $Al_xIn_{1-x}P$ (0.15<x<0.30) on $(Al_yGa_{1-y})_{0.52}In_{0.48}P$ (0.3<y<0.90) lattice-matched to GaAs led to spontaneous self-assembled nanostructures via Stranski–Krastanov growth.[40] Although 3 monolayers (ML) of InP on $(Al_{0.6}Ga_{0.4})_{0.52}In_{0.48}P$ deposited at 620 °C resulted in well-defined semi-spherical dots, typically emitting around ~700 nm (see figure S1 in SI), a significant blue-shifted emission can be achieved by Al incorporation and reduction of the number of monolayers down to 1-2. Figure 1a-b shows the surface morphology of a single uncapped layer of 0.4 nm (nominal) of $Al_{0.2}In_{0.8}P$ (grown on $(Al_{0.6}Ga_{0.4})_{0.52}In_{0.48}P$) on a 6° off GaAs substrate). The thin layer of material deposited, while developing elongated features on the surface, did not allow to distinguish between quantum dot- or dash-like structures or what could be a simple step bunched surface organization as typically found for MOVPE grown III-V alloys; we note that AlGaInP alloys grown by MOVPE tend to show only very short range surface organization, unlikely their arsenide and InP counterparts.[38, 41, 42]

Photoluminescence investigations of samples capped with 200 nm of $(Al_{0.6}Ga_{0.4})_{0.52}In_{0.48}P$ showed that the InP "dot-like" emission was shifted towards shorter wavelengths, around 600 nm. Figure 1c shows the PL spectra of three representative samples in which both the $Al_{0.2}In_{0.8}P$ (0.4 nm) and $(Al_{0.6}Ga_{0.4})_{0.52}In_{0.48}P$ layers were grown at the same temperature (620, 670 and 720 °C, respectively) compared to a reference sample where the $Al_{0.2}In_{0.8}P$ layers were grown at 620 °C and the $(Al_{0.6}Ga_{0.4})_{0.52}In_{0.48}P$ layers at 720 °C. The $(Al_{0.6}Ga_{0.4})_{0.52}In_{0.48}P$ emission showed a blue-shift with increasing growth temperature, as expected from the reduction of CuPt ordering[43] and also a reduction in intensity, possibly related to closer proximity to the direct-to-indirect bandgap transition. The increasing temperature also resulted in the $Al_{0.2}In_{0.8}P$ emission blue-shifting. The broad peak around 630 nm at 620 °C became a narrow emission around 590 nm, suggesting reduced lateral dimensions and possibly more uniform sizes of the $Al_{0.2}In_{0.8}P$ nanostructures.

The TEM cross section in figure 1d of a multilayer structure including 8 $Al_{0.2}In_{0.8}P/(Al_{0.6}Ga_{0.4})_{0.52}In_{0.48}P$ layers (AlGaInP barriers were 10 nm thick) highlights additional morphological details. The $Al_{0.2}In_{0.8}P$ layer appears to be not homogeneous in thickness, showing a tendency to accumulate in specific areas, with some hints of vertical coupling of fluctuations, although without a regular organization, either in-plane or out-of-plane. Nevertheless, the PL spectra acquired in samples with $Al_{0.2}In_{0.8}P/(Al_{0.6}Ga_{0.4})_{0.52}In_{0.48}P$ layer numbers ( $l$ ) ranging from 1 to 8 were all comparable in their emission characteristics (within our setup reproducibility), thus excluding significant coupling effects in the AlInP nanostructure morphology and between carrier wavefunctions in different layers (see figure S1 in SI).

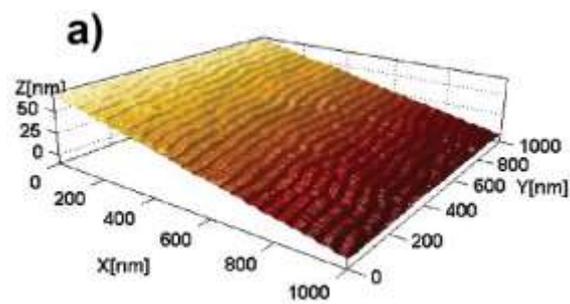

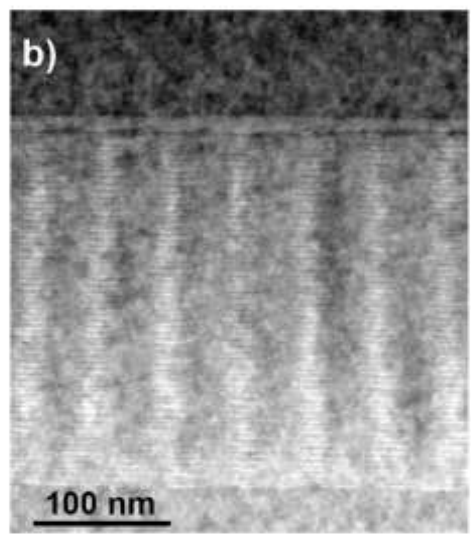

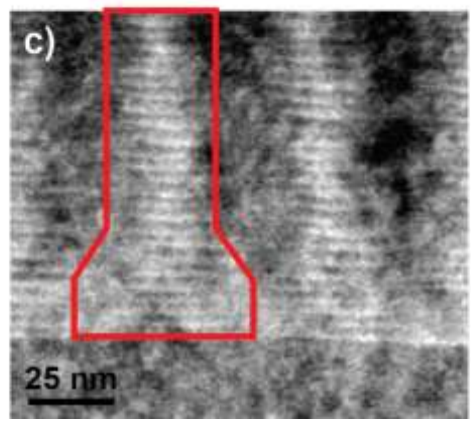

**Figure 2**: (a) AFM reconstructed 3D height image of an uncapped SMWR with 20 layers of 0.4 nm (nominal) of $Al_{0.2}In_{0.8}P$ separated by 3 nm (nominal) of $(Al_{0.6}Ga_{0.4})_{0.52}In_{0.48}P$ grown on a 6° off GaAs substrate. The surface appeared to have RMS below 1 nm and height variations in range of few hundreds of picometers along the wire. (b) Cross section TEM of a SMWR with 60 layers, including an $(Al_{0.5}Ga_{0.5})_{0.52}In_{0.48}P$ QW on top of the SMWR (wire direction is perpendicular to the plane). (c) Enlargement of the pedestal of a SMWR column.

The system morphology showed an interesting evolution with increasing number of layers $l$ when the $(Al_{0.6}Ga_{0.4})_{0.52}In_{0.48}P$ barrier was decreased down to 3 nm and $l$ increased to more than ~20. The surface morphology of the un-capped sample (figure 2a), where 20 $Al_{0.2}In_{0.8}P$ layers were stacked, and growth finished with the deposition of the last one (without capping layer), indicated the presence of wire-like ordering with length between 0.5-2 μm and width around 25 nm elongated orthogonally to the substrate tilt direction. The corresponding TEM cross section perpendicular to the wire long axis in figures 2b-c shows a highly ordered system with vertical columns of $Al_{0.2}In_{0.8}P$ wires having a periodicity of around 25 nm. Notably, the self-organization in columns shows a long-range ordering of microns length (along the substrate tilt direction), as demonstrated by the low magnification TEM and SEM pictures showing column arrays without any observable structural defect (see figure S2 in SI). Moreover, focusing at the bottom of each column, one can observe a pedestal-like structure at the beginning of the column formation process. Indeed, the first stacked layers are much more disordered, in agreement with the AFM scan of a single layer, as shown in figure 1a-b. From the cross-section TEM picture, it can be inferred that the system takes around 20 layers in order to self-organize in regular spaced columns arising from the vertical correlation of the in-plane wire formation. We speculate that this organization is fundamentally driven by strain forces related to the large lattice mismatch between the $(Al_{0.6}Ga_{0.4})_{0.52}In_{0.48}P$ layer, which is lattice-matched to GaAs, and the strained thin $Al_{0.2}In_{0.8}P$ layers.[40] It is also worth to note that after the required ~20 layers to reach an organized morphology, the resultant system is very stable for a further increase of $l$, allowing growth of samples with up to $l=200$. We also observed that multilayered structures resulted in very different morphologies on perfectly oriented GaAs substrates, and, in general, it was not possible to grow 20 or more layers without the appearance of evident structural relaxation and a resulting red-shift of the emission above 700 nm. Accordingly, we speculate that the surface step (bunched) organization in a 6° off GaAs substrate affects the $Al_{0.2}In_{0.8}P$ ad-atom incorporation and in-plane ad-atom mobility and constitutes a key factor for the long-range self-organization of the AlInP wires (see SI).

We finally observe that while the growth organization here seems unique to this material system and has no obvious equivalent in the literature, vertical nanostructure organization has been indeed reported several times in other systems. For early examples see references 44-46.

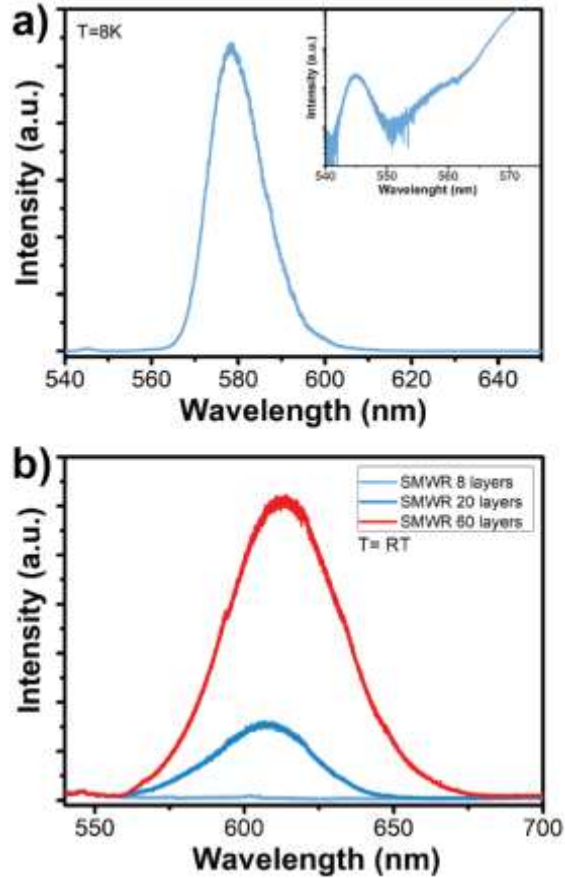

**Figure 3**: (a) 8 K photoluminescence spectrum of a SMWR with 60 layers, peaked at 580 nm. Insert highlights the weak $(Al_{0.6}Ga_{0.4})_{0.52}In_{0.48}P$ barrier emission around 545 nm. (b) Comparison between the photoluminescence spectra at room temperature of SMWRs with 8 (light blue line), 20 (blue line) and 60 (red line) layers. Notably, the 8 layer SMWR did not show any detectable emission at room temperature.

Figure 3a shows the photoluminescence spectrum at 8 K of a representative sample with $l$=60. The small peak around 545 nm can reasonably be attributed to the $(Al_{0.6}Ga_{0.4})_{0.52}In_{0.48}P$ barrier emission. The weak intensity of this peak was expected due to the indirect nature of the $(Al_{0.6}Ga_{0.4})_{0.52}In_{0.48}P$ bandgap ($(Al_xGa_{1-x})_{0.48}In_{0.52}P$ should show a direct-to-indirect transition at x~0.53). Interestingly, a further increase of the Al content, i.e. further increase of the indirect bandgap, decreased the $Al_{0.2}In_{0.8}P$ wire emission (see figure S3 in SI). Such a photoluminescence quenching was previously reported in AlGaAs and AlGaInP systems and attributed to an increase of point defects with the Al content in AlGaInP.[47] On the other hand, the stronger peak at 580 nm, tunable by varying the Al content in the AlInP compound, is attributed to the "wire" emission. Moving from 8 K to room temperature the wire emission was red-shifted by roughly 10-20 nm, comparable to the bandgap renormalization described by the Varshni equation, and dropped by more than 2 orders of magnitude in intensity.

Figure 3b compares the RT emission of three structures having 8, 20 and 60 $Al_{0.2}In_{0.8}P/(Al_{0.4}Ga_{0.6})_{0.52}In_{0.48}P$ layers respectively. Although the SMWR with $l$=8 did not show any detectable emission at RT, a drastic increase in intensity by at least two orders of magnitudes was observed when $l$ was increased from 8 to 20, while a linear trend with the layer number was found for $l$ >20. This is in agreement with the morphology evolution described in figure 2, in which around 20 periods were required for self-assembled ordering in the columnar nanowires to become established. We also noticed, since the morphology of a multilayer structure with $l$ >20 was very stable, that the room temperature intensity can be improved by a simple proportional increase of the number $l$ (this has been experimentally verified up to $l$ =200).

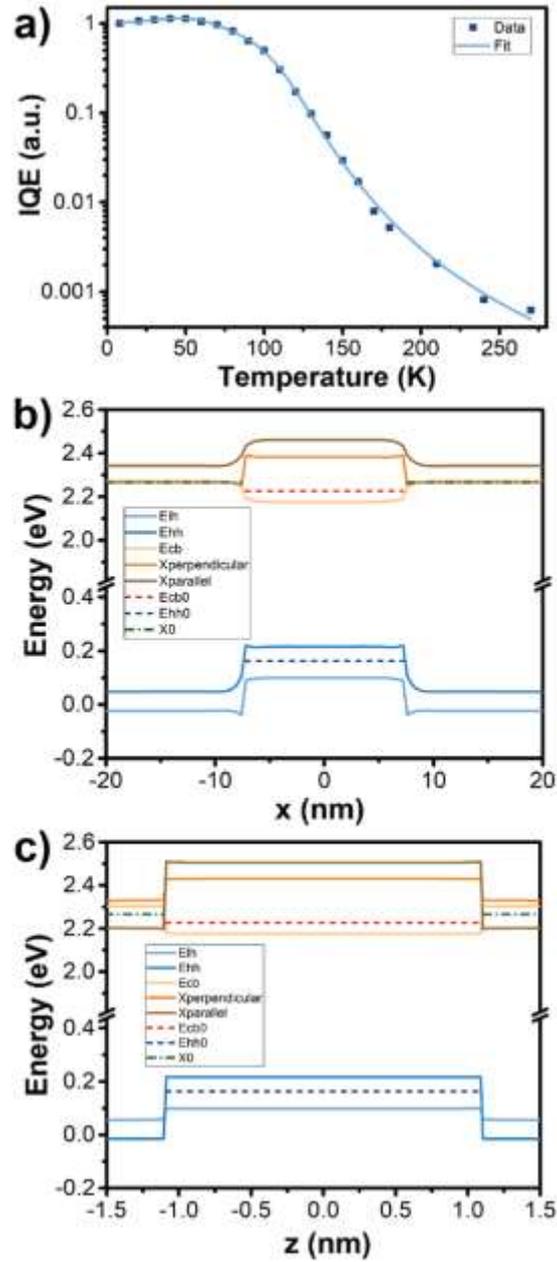

**Figure 4**: (a) Temperature dependent normalized integrated intensity. Blue line indicates the data fit obtained with eq. 1. (b) Theoretical simulation of the expected band alignment in the growth plane in a 15 nm thick and 2.2 nm tall $Al_{0.2}In_{0.8}P$ wire surrounded by $(Al_{0.6}Ga_{0.4})_{0.52}In_{0.48}P$ at 300 K, plotted along a line through the center of the wire, and perpendicular to the wire axis, and (c) plotted through the center of the wire along the growth direction. The solid lines depict the bulk band edges, and the dashed lines correspond to the ground confined states. The vertical and horizontal separation between wires is 1.2 nm and 25

nm, respectively.

The evolution of the IQE with temperature of a SMWR with *l*=60 is presented in figure 4a. The overall trend expected for an AlGaInP system is a decrease in emission intensity with increasing temperature, attributed to thermally activated escape of excitons from the (poorly) confined regions (wires here, wells in conventional red emitters) to the barriers and to subsequently non-radiative recombination processes.[47,48] However, we observe that before the drop in intensity for T>100 K the integrated intensity showed an anomalous increase, reaching its maximum around 50-60 K.[49] Moreover, a model based on a single escape process was not able to reproduce the data trend above that temperature (see SI).[48]

In order to understand this behavior and get more insights into the system dynamics, we simulated the expected band structure of the $Al_{0.2}In_{0.8}P/(Al_{0.6}Ga_{0.4})_{0.52}In_{0.48}P$ SMWR system (figure 4b-c), using an 8-band **k·p** model for the zone-center Γ states, a one-band model for each of the three X states, and ignoring the effects of substrate tilt on the calculated states. Details of the wire structure considered, and of further variations on wire structure are given in SI. The conduction band minimum in the $(Al_{0.6}Ga_{0.4})_{0.52}In_{0.48}P$ barrier is located, as expected, in the X valley. The 2.2% lattice mismatch between the $Al_{0.2}In_{0.8}P$ SMWR and the GaAs substrate results in a strain distribution across the SMWR and the surrounding barrier layers, causing a further reduction of the indirect band gap in the barrier by splitting the X states into $X_\perp$ and $X_\parallel$ (see SI), where $X_\perp$ are the two X states in the growth plane, while $X_\parallel$ is the X state along the growth direction. The band alignment for *l*=60 in figure 4b-c shows that the X states in the quantum wires act as a potential barrier, with the lowest X states in the barrier material. For the SMWR structure considered here, the lowest confined X state is in the $X_\perp$ valley, although it is possible, e.g. by assuming a wider vertical separation between wires, to have the lowest confined X state in the $X_\parallel$ valley. In either case, however, the X states have larger electron effective mass compared to the Γ valley. Assuming the transition of the Γ-like electrons in the quantum wire into the barrier X valley is the likely thermally activated non-radiative pathway in this structure, we find that the potential barrier for direct gap electrons is 10 – 40 meV, depending on exact structure assumed (40 meV in figure 4b-c). On the other hand, heavy holes are confined much deeper in the wire, with the potential barrier between 120 – 195 meV (53 and 178 meV in figure 4b-c respectively to the barrier heavy-hole band edge). Our calculations also show that the variation of wire geometry, which can be seen in figure 2, does not change the (ground state) transition energy significantly enough to reproduce the width of the PL spectra, and it is likely that interdiffusion of Al and Ga occurs between the wire and barrier during growth (see SI).

Building on this analysis, we have used a simple model to describe the temperature dependence of the emission intensity observed in figure 4a. Since the conduction band minimum in AlGaInP is at the X point in reciprocal space, the carrier migration rate from the barriers into the wires is reduced at low temperature, due to the low number of phonons available. However, the electron-phonon scattering probability increases approximately quadratically with temperature, thus allowing excited carriers in the barriers to diffuse into the wires and enhance the wire emission intensity. For temperatures higher than 50-60 K, confined carriers can leak out from the wire. In particular, electrons should be the first to have enough thermal energy to escape, based on the asymmetric carrier confinement energies derived from theory. This process should be followed by a similar thermally activated emission of holes at higher temperatures. If so, the thermally activated migration of carriers from the wire to the barriers and the availability of non-radiative recombination paths outside the wire should be considered the dominant mechanism behind the loss of intensity at room temperature.

According to this, we can estimate the formal behavior of temperature dependency of the integrated intensity *I(T)* as (see SI)

$$I(T) \propto \frac{1}{1 + \tau_r(T)\left(\frac{1}{\tau_{es}^e(T)} + \frac{1}{\tau_{es}^h(T)} - \frac{1}{\tau_{in}(T)}\right)}$$

where $\tau_r$ is the radiative lifetime, $\tau_{es}^e$ and $\tau_{es}^h$ the escape time constant for electron and holes respectively and $\tau_{in}$ the time constant for carrier injection from the barriers to the wires. The variation of the radiative lifetime with temperature can be described as $\tau_r(T) \sim \sqrt{T}$ for a wire ($\tau_r(T) \sim T$ in a well), with $\tau_{es}^{e/h}(T) \sim e^{E_{act}/kT}$ and $\tau_{in}(T) \sim T^{-2}$.[50, 51]

The data fit showed in figure 4a reproduces well the data across the whole range of temperature. In particular the fit provides ~20 meV and ~120 meV as activation energies for electrons and holes respectively, in reasonable agreement with the theory. It should be noted that according to theoretical predictions electron confinement in $(Al_{0.3}Ga_{0.7})_{0.52}In_{0.48}P$ QWs (emitting at the same wavelength) is expected to be larger, potentially leading to a higher IQE at high temperatures.

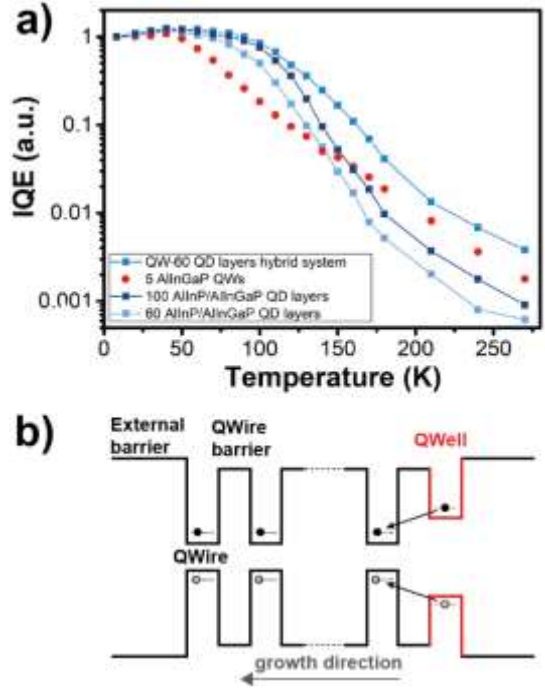

**Figure 5**: (a) Comparison between the temperature dependent normalized integrated intensities of a SMWR with 60 layers (light blue), a SMWR with 100 layers (blue), a QW-SMWR with 60 layers (dark blue) and a 5 QW $(Al_{0.3}Ga_{0.7})_{0.52}In_{0.48}P/(Al_{0.4}Ga_{0.6})_{0.52}In_{0.48}P$ structure (red). (b) Schematic illustration of the band alignment in the hybrid QW-SMWR system in which the QW is used to inject carriers into the SMWR system.

In order to investigate this aspect, figure 5a compares the temperature-dependent IQE of SMWRs with $l$=60 and $l$=100, respectively, and a 5QW $(Al_{0.3}Ga_{0.7})_{0.52}In_{0.48}P/(Al_{0.4}Ga_{0.6})_{0.52}In_{0.48}P$ structure emitting at similar wavelength (the number of QWs and their thickness has been designed to sum up to give approximately the same amount of active material deposited as in our wire structures with 100 layers, i.e. with a comparable overall nominal thickness of the emitting layers). We assume the ~0 K IQE to be near unity as is custom in these cases. While for T<150 K the SMWR IQE is larger than the that of the QW structure up to 1 order of magnitude (at T=100 K), at RT the QW structure IQE exceeds that of the beats SMWRs by more than a factor of 2.

However, a much larger improvement in the room-temperature efficiency of our structures can be obtained in a hybrid system in which a single quantum well is coupled to the SMWR system. In this scheme of work, depicted in figure 5b, the additional $(Al_{0.5}Ga_{0.5})_{0.52}In_{0.48}P$ quantum well, having a bandgap slightly larger than the wires (i.e. not contributing to the wire emission photoluminescence), can be used as a reservoir to inject carriers into the wires, and partially compensate the thermally activated quenching of the emission.[52] The IQE of the resultant hybrid device, containing just 60 layers (just over half that of the SMWRs shown by the dark blue dots in figure 5a), showed more than one order of magnitude higher efficiency compared to the system without a QW, and exceeds the efficiency of the 5QW $(Al_{0.3}Ga_{0.7})_{0.52}In_{0.48}P/(Al_{0.6}Ga_{0.4})_{0.52}In_{0.48}P$ structure over the full temperature range. We also noticed that a minor improvement in the overall efficiency was obtained by increasing Al content in AlGaInP barriers before and after the multilayer structure from $(Al_{0.6}Ga_{0.2})_{0.52}In_{0.48}P$ to $(Al_{0.8}Ga_{0.2})_{0.52}In_{0.48}P$ and by adding some strain to these barriers (see SI). This result demonstrates that the SMWR structure can be engineered not only to achieve a specific target wavelength, but also to improve red emitter efficiency by structural design.

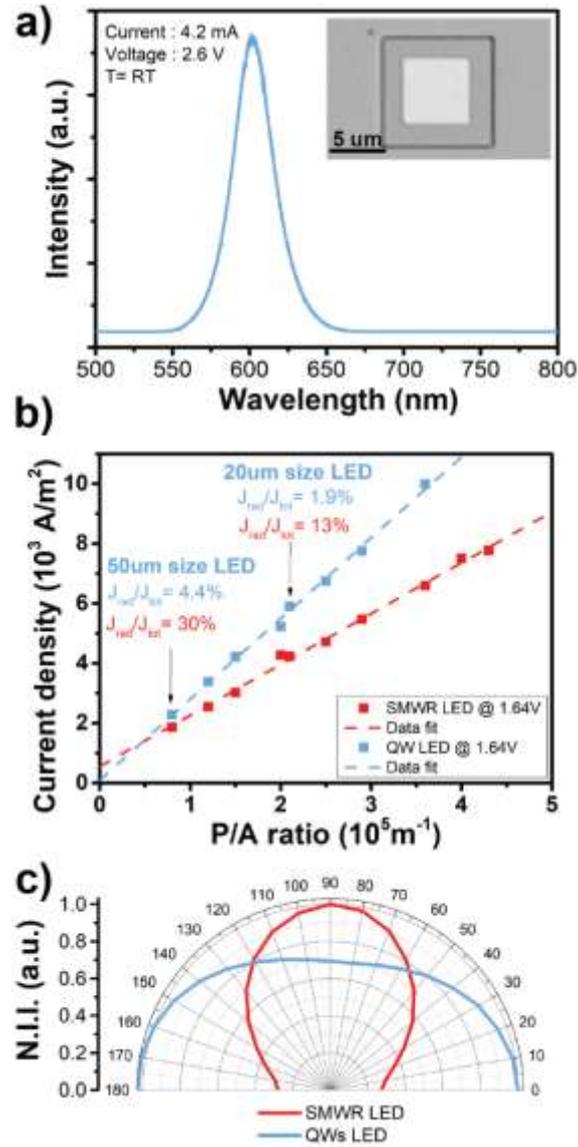

**Figure 6**: (a) Electroluminescence spectrum of a representative SMWR LED. Inset shows an optical image of the measured device. (b) Plot of current density versus Perimeter/Area (P/A) ratio, demonstrating a larger contribution of the surface current to the total current in a 5QW $(Al_{0.3}Ga_{0.7})_{0.52}In_{0.48}P/(Al_{0.6}Ga_{0.4})_{0.52}In_{0.48}P$ LED then in a SMWR LED. No significant spectral shift was observed in these current intervals (c) A normalized polar plot of the integrated intensity of linearly polarized light as a function of analyzer's angle. The light was collected from top-view from both LED devices.

To test the feasibility of higher efficiency light emitting electrical devices, a prototype SMWR microLED device (structure details in SI), was compared to a similarly processed 5QW $(Al_{0.3}Ga_{0.7})_{0.52}In_{0.48}P/(Al_{0.6}Ga_{0.4})_{0.52}In_{0.48}P$ LED. A representative electroluminescence spectrum of a SMWR LED device indicated an optical emission peaked at 610 nm (at room temperature), in agreement with the photoluminescence spectrum of the equivalent SMWR structure and showed a turn on voltage of ~1.6 V corresponding to a current density of ~0.1 A/cm$^2$. Similar values were observed for the QW LED, indicating that the low current density required should be ascribed to the current-spreading layer design rather than the material structure. Investigation of LED devices with comparable surface area but different perimeters unveiled important differences in the surface current contribution for the two types of LEDs. In a first approximation, the total current density injected into the LED can be described by[53]

$$J_{TOT} = J_{bulk} + \sigma_{surf}\frac{P}{A}$$

where $J_{bulk}$ is the current density flowing through the bulk of the material, $\sigma_{surf}$ (A/m) the surface current per unit length of perimeter P and A the area of the LED, respectively. Assuming that the carriers flowing through the surface do not recombine radiatively, and the emission intensity is only proportional to $J_{bulk}$, the percentage of the injected current $J_{bulk}/J_{TOT}$ that contributes to the radiative emission can be inferred from the linear fit of the data. In particular, two different trends for the QWs and the SMWRs LEDs were observed. In both devices only a small fraction of the injected current was actually passing through the bulk of the material. In the 50x50 μm$^2$ QW LED only 4.4 % of the total current is contributing to the emission, with this value decreasing to 1.9 % in a 20x20 μm$^2$ device. On the other hand, the value of $J_{bulk}/J_{TOT}$ in the SMWR LED was found to be 30 % and 13 % for the 50x50 μm$^2$ and 20x20 μm$^2$ device, respectively, almost one order of magnitude larger. The reduction of the surface leakage current was further supported by a reduction in the current spreading (see SI) in the SMWR LED. In fact, the higher degree of space confinement and potential fluctuations experienced by carriers in the wires, compared to that in the QWs, reduces the carrier diffusion towards the LED surface before recombination. This therefore suggests that higher current-to-light conversion efficiency and a LED size reduction without any loss in efficiency are achievable in QW-SMWRs.

An important additional benefit of SMWR LEDs is that their spontaneous light emission is polarized predominantly along the wire main axis. Figure 6c shows the normalized integrated intensity (NII) of linearly polarized light (filtered for polarization as described in SI) versus the sample rotation angle for the SMWR and QW LEDs. The QW LED showed only a weak sign (around 30 %, for definition see SI) of polarization along the substrate tilt direction, indicating that the tilted GaAs substrate might affect the QW morphology or introduce residual strain, resulting in a low degree of polarization. On the other hand, the wires possessed strong emission anisotropy, with almost 80 % of the emitted light polarized along the wire main axis, i.e. perpendicular to the substrate tilt direction, making them extremely interesting for modern display/augmented reality (and similar) applications. It should be noted also that the presence of a large metal contact covering more than 50 % of the LED top surface reduced the amount of direct light that could be collected by the optical setup; therefore we expect that the degree of polarization could have been reduced in our measurements by scattered light from the sidewalls of the LED.

In conclusion, we presented a novel multilayer structure comprising self-assembled columns of AlInP wires with tunable emission between 580 and 650 nm. The growth condition and morphology was investigated and optimized, resulting in a reliable method to growth a large number of $Al_{0.2}In_{0.8}P/(Al_{0.6}Ga_{0.4})_{0.52}In_{0.48}P$ layers at high temperature (720 °C). The structure design was engineered to maximize the efficiency by optimizing AlInP and AlGaInP layer thicknesses, growth temperatures and Al content in the barriers. A qualitative physical model based on theoretical simulations was proposed and demonstrated capable to describe satisfactorily the SMWR temperature-dependent photoluminescence. The IQE was investigated and compared to a 5QW AlGaInP sample and to a QW-SMWR sample, demonstrating that adding a QW coupled to the SMWR system boosts by more than one order of magnitude the IQE, outperforming the AlGaInP QWs. Finally, investigation of the QW-SMWR microLED prototype indicated further advantages over the traditional QW LED in the reduced surface parasitic current and in the emission of strongly polarized light.

These results provide an interesting perspective for SMWR structures for polarized high-efficiency yellow/amber/red light emitting diodes.

AUTHOR INFORMATION

**Corresponding Author**

*agnieszka.gocalinska@tyndall.ie

**Author Contributions**

The manuscript was written through contributions of all authors. All authors have given approval to the final version of the manuscript.

SUPPORTING INFORMATION

Additional details and Supplementary Figures S1-S6.

ACKNOWLEDGMENT

This research was enabled by Science Foundation Ireland under the IPIC award 12/RC/2276, grant 10/IN.1/I3000, 15/IA/2864 and the Irish Research Council under grant EPSPG/2014/35. We thank Kevin Thomas for the MOVPE support.

# Supporting Information for: Three-dimensional Self-assembled Columnar Arrays of AlInP Quantum Wires for Polarized Micron-sized Amber Light Emitting Diodes

*Andrea Pescaglini,[a] Agnieszka Gocalinska,[a,]* Silviu Bogusevschi,[a,b] Stefano Moroni,[a] Gediminas Juska,[a] Enrica Mura,[a] John Justice,[a] Brian Corbett,[a] Eoin O'Reilly,[a,b] Emanuele Pelucchi[a]*

[a] Tyndall National Institute – University College Cork, Lee Maltings, Cork, Ireland

[b] Department of Physics, University College Cork, Ireland

SUPPORTING INFORMATION

1. **Sample characterization**

The detailed morphological study was performed with atomic force microscopy (AFM) in tapping/noncontact mode at room temperature and in air. Layer composition was determined by x-ray diffraction measurements. The temperature dependent photoluminescence measurements were carried out in a closed-cycle cryostat in a conventional micro-photoluminescence set-up with confocal arrangement. The samples were excited non-resonantly with a single-mode, continuous-wave semiconductor laser diode emitting at 532 nm. Light was collected with a long working distance 50x objective with NA=0.5, dispersed by a 950 grooves/mm diffraction gratings, and intensity measured with a nitrogen-cooled CCD. Electroluminescence of the devices was measured in the equivalent optical set-up, however, equipped with a probe station allowing contacting the devices electrically during the measurement procedure. The light was collected by a 10x, long-working-distance objective. Polarization was analyzed by placing a broadband half-wave-plate ($\lambda/2$) mounted on a motorized rotator and a polarizer, both stacked above the objective to avoid polarization state distortion by other elements present in the optical path. The quoted polarization percentage refers to: $(I_{max}-I_{min})/I_{max}$, with the maximum/minimum signal intensity across perpendicular directions.

2.  **LED fabrication**

P-metal contact regions are defined by lift-off lithography on the top surface of the LED wafers.  A surface de-oxidation step is applied by 10 second immersion of samples in Buffered Oxide Etch (BOE) diluted with deionised water by volume 1:5. P-type metal consisting of Ti/Pt/Au (20/30/200 nm) is electron beam evaporated onto the surface. A mask of 800 nm Silicon Dioxide is sputtered onto the surface.  Photoresist mesas are defined over the P-contacts.  The hard mask is then dry etched by an Inductively Coupled Plasma (ICP) using a mixture of $CF_4/CHF_3$. Photoresist is removed by a solvent rinse.  The LED material is dry etched by ICP using a mixture of $BCl_3/Cl_2/Ar$ at 150°C or alternately $SiCl_4/He$ at 80°C through the epitaxial layers to the n-substrate. End-Point Detection (EPD) is used to ensure etching has been carried out to the correct depth. The $SiO_2$ hard mask is removed using the same etch as used for defining the mask. An N-metal of Au/Ge/Au/Ni/Au is applied to the clean substrate side of the wafers.  Samples are then ready for testing. A furnace anneal of 420°C for 5 minutes in a $N_2/H_2$ (95:5) atmosphere is applied to anneal the contacts.

## 3. Morphology characterization

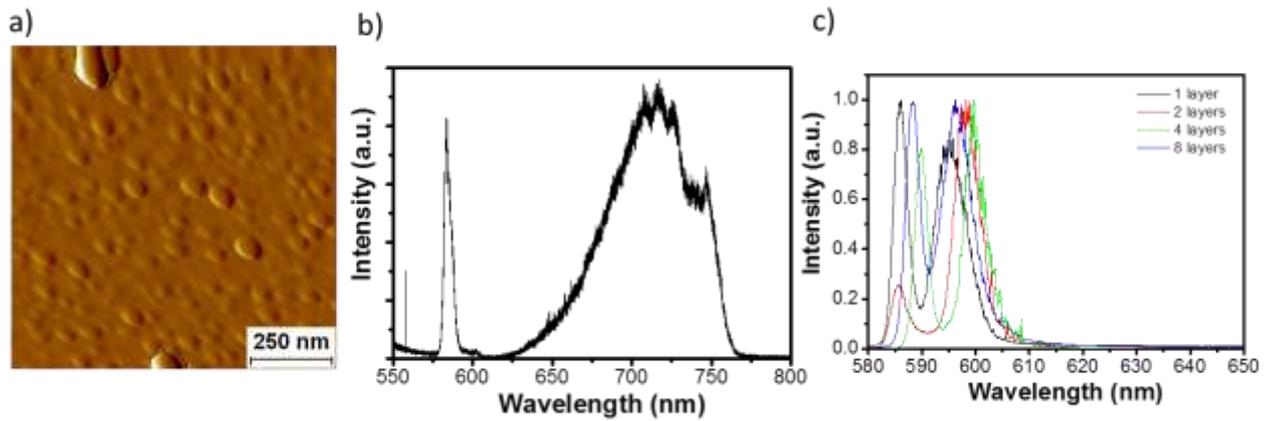

**Figure S1**: (a) AFM image of 0.5 nm (nominal) InP on $(Al_{0.3}Ga_{0.7})_{0.52}In_{0.48}P$ on 6° off GaAs substrate. (b) Room temperature photoluminescence of the sample showed in (a) capped with 200 nm of $(Al_{0.3}Ga_{0.7})_{0.52}In_{0.48}P$. (c) Comparison between the photoluminescence spectra of 0.5 nm (nominal) $Al_{0.2}In_{0.8}P$ separated by 10 nm of $(Al_{0.3}Ga_{0.7})_{0.52}In_{0.48}P$ for an increasing number of layers from 1 to 8.

## 4. Long range ordering of SMWR

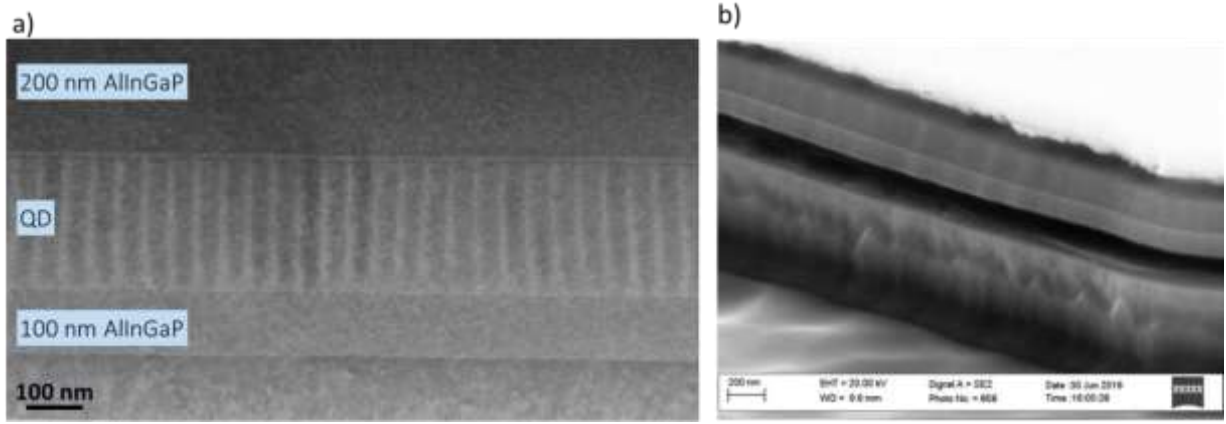

**Figure S2**: (a) Low magnification cross-section TEM image of a QW-SMWR with 60 layers perpendicular to the wire long axis. The $(Al_{0.5}Ga_{0.5})_{0.52}In_{0.48}P$ QW is on top of the SMWR. No defects were observed. (b) SEM side-view image of a SMWR LED showing the columnar organization on a micron scale range without any visible morphological defect.

## 5. Effect of the Al content in AlInGaP barriers on the SMWRs emission intensity and temperature dependence photoluminescence

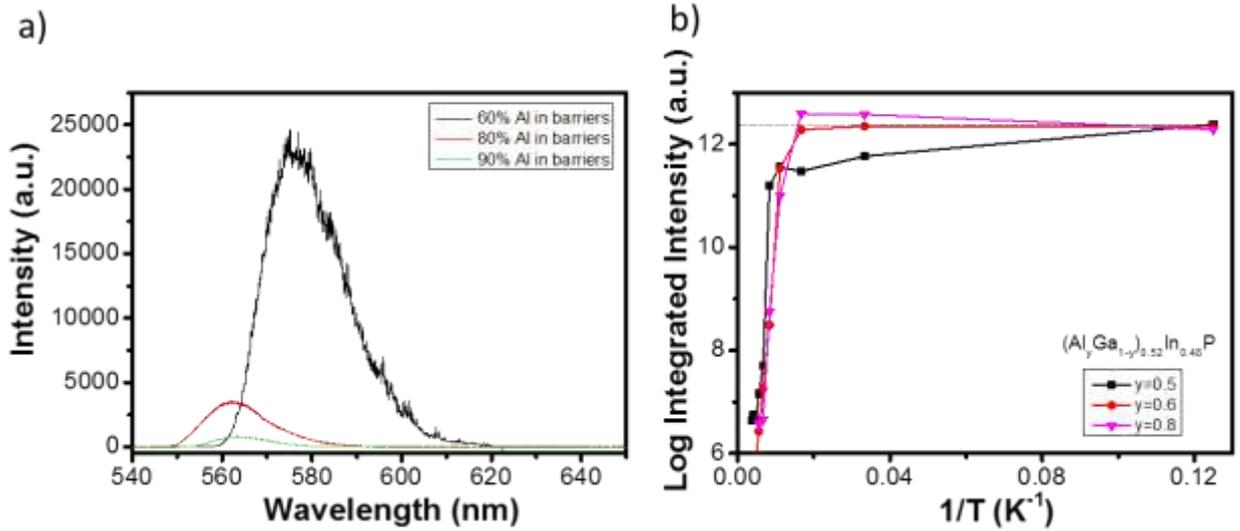

**Figure S3**: (a) Photoluminescence spectra of three $Al_{0.2}In_{0.8}P/(Al_yGa_{1-y})_{0.52}In_{0.48}P$ SMWRs with 8 layers, 10 nm barriers and $0.6<y<0.9$ (b) Temperature dependent integrated intensity of 8 layers SMWR with different wire and barrier compositions. The increase in integrated intensity above the value at 8 K (crossing the black dotted line) appeared when $(Al_yGa_{1-y})_{0.52}In_{0.48}P$ barrier has $y>0.5$.

The relative Al/Ga content in the AlInGaP barriers has been investigated to maximize the light emission while remaining lattice-matched to GaAs. Contrary to initial expectations, increasing the Al content above the direct-to-indirect transition point did not improve the emission intensity despite that the larger Al content should increase the indirect bandgap of the barriers. Figure S3a compares the PL emission of three samples with nominal Al content in the barrier 60 %, 80 % and 90 %. Although we did observe an increase of PL emission when the Al content was increased from 30 % to 60 % (not shown), a drop in intensity is instead evident for Al content above 60 % (a small blue-shift of the emission, attributed to a possible diffusion of Al from the barriers to the dots, is also evident). A stronger photoluminescence quenching with increasing Al content in AlGaAs and AlInGaP has been previously reported (although, probably not well understood) and attributed to point defects.[4,5]

Figure S3b shows the temperature-dependent integrated intensity for 60 layers SMWRs with relative Al content ranging from 50 to 80 %. The anomalous increase in integrated intensity appeared for Al>50 % suggesting a correlation with the indirect bandgap, in agreement with the model described in the article.

## 6. Details of the 8-band k·p model

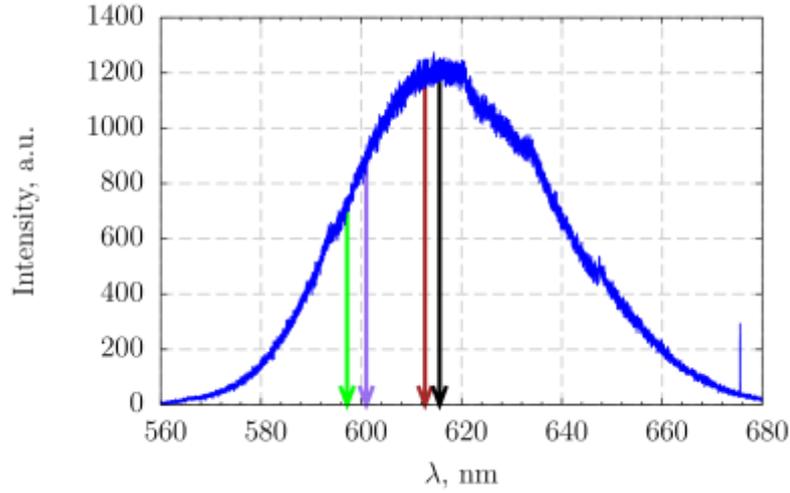

**Figure S4**: Photoluminescence spectrum of $Al_{0.2}In_{0.8}P/(Al_{0.6}Ga_{0.4})_{0.52}In_{0.48}P$ SMWR with 60 layers at 300 K. The vertical arrows show the calculated transition energies for SMWR with different geometries and compositions (for details see Table S1).

| Line color | QWire | | | | | Barrier | $\lambda$, nm |
| --- | --- | --- | --- | --- | --- | --- | --- |
| | Composition | $d_w$, nm | $h_w$, nm | $d_{xy}$, nm | $d_z$, nm | Composition | |
| Green | $Al_{0.22}In_{0.78}P$ | 7.5 | 2.2 | 32.5 | 1.8 | $Al_{0.32}Ga_{0.20}In_{0.48}P$ | **597.2** |
| Purple | $Al_{0.22}In_{0.78}P$ | 15 | 2.2 | 25 | 1.8 | $Al_{0.32}Ga_{0.20}In_{0.48}P$ | **601.0** |
| Brown | $Al_{0.177}Ga_{0.043}In_{0.78}P$ | 15 | 2.2 | 25 | 1.8 | $Al_{0.32}Ga_{0.20}In_{0.48}P$ | **612.7** |
| Black | $Al_{0.135}Ga_{0.085}In_{0.78}P$ | 7.5 | 2.2 | 32.5 | 1.8 | $Al_{0.33}Ga_{0.1}In_{0.48}P$ | **615.6** |

**Table S1**: Composition and geometry data used for theoretical simulations of the SMWR system, with $d_w$, $h_w$, $d_{xy}$ and $d_z$ corresponding to the QWire width, QWire height, in-plane and vertical distance between the QWires respectively. The calculated transition energies in units of wavelength $\lambda$ correspond to the highlighted arrows in the Figure S4.

Theoretical modeling is used to (a) estimate transition energies and band offsets in the SMWR structures and (b) to analyze the temperature dependence of the PL emission intensity. We ignore the effects of substrate tilt, assuming growth along the (001) direction, and using an 8-band **k·p** model for the zone-center Γ states and a one-band model for each of the three X states. Strain effects are calculated using second order elasticity theory. For the initial calculation (shown in figure 4 (b) and (c)), we assume no interdiffusion between the dot and barrier material, and that the ratio of dot to barrier material follows the nominal layers' thicknesses of 0.4 nm (dot) and 3 nm (barrier). Based on the upper layers in figure 2(b) and figure S2 (a), we first assume rectangular wires of height 2.2 nm and width 7.5 nm, with vertical and in-plane repeat distance of 3.4 nm and 40 nm respectively. This gives emission (green arrow in figure S4) on the short wavelength side of the measured 300 K PL spectrum for the structure being considered. Table 1 shows the emission wavelength calculated for a range of further changes in dot and barrier composition, dimensions and periodicity. If we double the wire base length (breaking the assumed nominal composition ratio of 0.4:3.0), this leads to a small red shift in the emission wavelength, but still on the short wavelength side of the measured PL spectrum as indicated by the purple arrow in figure S4. One route to longer wavelength emission is to allow interdiffusion of Al and Ga between the wire and barrier regions, as illustrated by the brown and black arrows in figure S4, Overall, we conclude that the band offsets and confinement energies presented in figure 4 (a) and (b) provide a good starting point to discuss recombination in the SMWRs investigated.

### 7. Phonon-electron injection model

The temperature-dependent integrated intensity of SMWRs showed an anomalous increase of emission intensity at low temperatures with a maximum peak around 40-50 K. This feature, scarcely reported in literature, was reproducibly found in all samples with barriers' Al content higher than 50%, while no evidence was found in samples with direct bandgap barriers (see fig S3b). Since this phenomenon was correlated to the direct-to-indirect bandgap transition of the AlInGaP barriers, we speculate that an electron-phonon scattering process might be responsible for the increase in the emission intensity. Indeed, at low temperature, where the carriers are well confined at the Γ point of the wire and the thermal-activated escaping can be neglected, phonon-scattering events may let electrons/holes diffuse from the barriers to the wires, including the intraband transitions from the X point to the Γ point for electrons, where they remain confined and eventually recombine radiatively. In particular, we can speculate that the recombination lifetime at the X point in the barriers is probably longer than the radiative recombination lifetime at Γ, due to both the k-mismatch as well as a potentially larger effective mass leading to a reduced mobility. Therefore electron-hole pairs excited by the laser radiation (with energy above the barrier bandgap) have enough time to move from the barrier into the wire via scattering with the phonons before recombination. This diffusion process, mainly activated by the interaction with the lattice (i.e. by temperature), can then lead to an increase in the integrated intensity with the temperature (below 80 K). By contrast, this process may be less likely in barriers with a direct bandgap, where the carrier lifetime may be too short to allow diffusion from the barriers to the wires.

For a further rise in temperature however, the electrons and holes have increasing probability to escape, followed by carrier redistribution across the entire superlattice conduction (valence) band and non-radiative recombination resulting in a photoluminescence quenching. However, according to the theoretical band alignment discussed in the article, electrons and holes have different confinement energies. Therefore, we propose a simple model that takes into account the phonon-mediated injection of carriers into the wire and the thermally activated escape of electrons and holes with different activation energies.

In the simple model of a system with two thermally activated non-radiative recombination mechanisms, the relation between $\tau_g$ the carrier generation rate, $\tau_r$ the radiative recombination rate, and $\tau_{es}^e$ and $\tau_{es}^h$ the two escapement rates is, following refs [5, 6]

$$\frac{1}{\tau_g} = \frac{1}{\tau_r} + \frac{1}{\tau_{es}^e} + \frac{1}{\tau_{es}^h}$$

Assuming an additional electron-phonon injection mechanism with a scattering rate $\tau_{in}$ the relation becomes, with this ansatz

$$\frac{1}{\tau_g} = \frac{1}{\tau_r} + \frac{1}{\tau_{es}^e} + \frac{1}{\tau_{es}^h} - \frac{1}{\tau_{in}}$$

Since the emission intensity as function of the temperature I(T) is proportional to the ratio

$$I(T) \propto \frac{1/\tau_r}{1/\tau_g}$$

We obtain by substitution the final expression (always assuming this as a reasonable approximation)

$$I(T) \propto \frac{1}{1 + \tau_r(T)\left(\frac{1}{\tau_{es}^e(T)} + \frac{1}{\tau_{es}^h(T)} - \frac{1}{\tau_{in}(T)}\right)}$$

Here the temperature dependence of the radiative lifetimes with the temperature can be described as $\tau_r(T) \sim \sqrt{T}$ for a wire $\tau_r(T) \sim T$ in a well), $\tau_{es}^{e/h}(T) \sim e^{E/kT}$ and $\tau_{in}(T) \sim T^{-2}$. The experimental data were fitted with the following function

$$I(T) \propto \frac{1}{1 + \sqrt{T}(Ae^{-\frac{E_1}{kT}} + Be^{-\frac{E_2}{kT}} - CT^2)}$$

with A,B,C and the escape energies $E_1$ and $E_2$ free parameters of the model. As shown in figure 4a this simple extension of the thermally activate non-radiative recombination model can reasonably fit the data over the entire temperature range. The estimated activation energies were around 20 meV and 120 meV for the SMWR samples while in the QW sample they were 20 and 80 meV. Interestingly the low activation energy value and the coefficient related to the electron-phonon scattering term are very comparable in all the structures considered. This suggests that are independent of the 2D or 1D geometry used in the different samples. We tentatively attribute the two energies to the electron confinement and hole confinement energies respectively. In this picture then, the larger carrier confinement in wires compared to the well induced by the reduced dimensionality resulted fundamentally in a larger confinement for holes only. We note that this is a comparison between an AlInP QWire and AlGaInP QWell (that have relatively similar ground state transition energy). Thus the difference in confinement between the two structures can be mainly caused by different band alignment (the electron and hole activation energy in the QWire (QWell) are ~24 and ~120 (103 and 53) meV respectively). At the same time, by reducing the dimensionality of the structure it may not necessarily mean that the hole confinement is improved. Different wire geometry, intrinsic to the Stranski-Krastanov growth, can result in a range of confinement energies. The primary effect of the reduced structure dimensionality typically leads to a shallower confinement.

We should point out that the two activation energies could be attributed also to two exciton confinements in different parts of the multilayer structure. In fact, figure 2c shows that in the "pedestal" of the column the wires are not well defined and the system is much more disordered. This in turn may lead to a lower carrier confinement in this region. However, the fact that this two activation energy model is also required for the QW system, giving a similar value of electron confinement, supports our former description rather than the latter.

**Further optimization of SMWR structures**

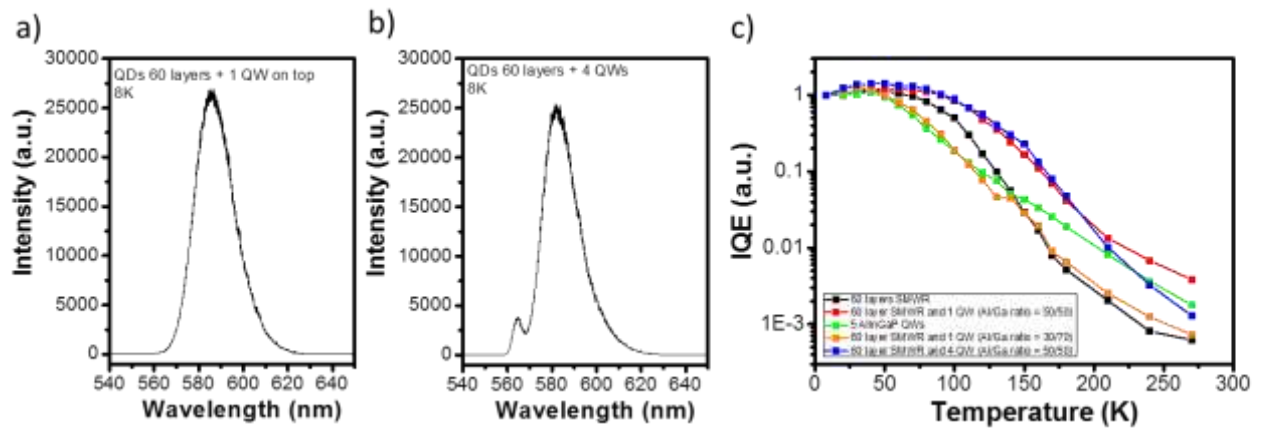

**Figure S5**: Photoluminescence spectra of a $Al_{0.2}In_{0.8}P/(Al_{0.6}Ga_{0.4})_{0.52}In_{0.48}P$ a SMWR structure with 60 layers coupled to (a) single $(Al_{0.5}Ga_{0.5})_{0.52}In_{0.48}P$ QW on top and (b) 4 $(Al_{0.5}Ga_{0.5})_{0.52}In_{0.48}P$ QWs inserted every 20 $Al_{0.2}In_{0.8}P/(Al_{0.6}Ga_{0.4})_{0.52}In_{0.48}P$ layers. (c) Temperature dependent IQE of a $Al_{0.2}In_{0.8}P/(Al_{0.6}Ga_{0.4})_{0.52}In_{0.48}P$ SMWRs structure with 60 layers and similar SMWR structures having a single $(Al_{0.3}Ga_{0.7})_{0.52}In_{0.48}P$ QW on top, single $(Al_{0.5}Ga_{0.5})_{0.52}In_{0.48}P$ QW on top and four $(Al_{0.5}Ga_{0.5})_{0.52}In_{0.48}P$ QWs placed every 20 layers of the SMWR structure. For comparison also the temperature dependence of a $(Al_{0.3}Ga_{0.7})_{0.52}In_{0.48}P/(Al_{0.6}Ga_{0.4})_{0.52}In_{0.48}P$ 5QW structure is also presented.

The optimization of the SMNW structure included careful considerations on the top and bottom AlInGaP barriers confining the SMNW multilayers and the position of the coupled QW. Figure S5 compares the IQE of a $Al_{0.2}In_{0.8}P/(Al_{0.6}Ga_{0.4})_{0.52}In_{0.48}P$ SMWR structure with 60 layers (wire emission at 580 nm at 8 K) with similar SMWR structures having a single $(Al_{0.3}Ga_{0.7})_{0.52}In_{0.48}P$ QW emitting at 590 nm (at 8 K) on top, a single $(Al_{0.5}Ga_{0.5})_{0.52}In_{0.48}P$ QW emitting at 560 nm (at 8 K) on top and four $(Al_{0.5}Ga_{0.5})_{0.52}In_{0.48}P$ QWs emitting at 560 nm (at 8 K) placed every 20 layers of the SMWR structure. For comparison the temperature dependence of a $(Al_{0.3}Ga_{0.7})_{0.52}In_{0.48}P/(Al_{0.6}Ga_{0.4})_{0.52}In_{0.48}P$ 5QW structure is also reported.

We observed that the SMWR with a $(Al_{0.3}Ga_{0.7})_{0.52}In_{0.48}P$ QW behaves differently from the $Al_{0.2}In_{0.8}P/(Al_{0.6}Ga_{0.4})_{0.52}In_{0.48}P$ SMWR structure with 60 layers and resembles the trend of the 5QW $(Al_{0.3}Ga_{0.7})_{0.52}In_{0.48}P /(Al_{0.6}Ga_{0.4})_{0.52}In_{0.48}P$ structure. This result suggest that when a QW with smaller bandgap than the $Al_{0.2}In_{0.8}P$ wires is coupled to the system, carriers localized in the wire at high temperatures migrate to the QW, resulting in an emission entirely dominated by the QW. On the other hand, a SMWR with 1 and 4 QWs having a smaller bandgap with respect to the wires showed higher efficiency at all temperatures. In particular, we noticed that the emission of a single QW was not observed, and only a weak emission was obtained by 4 QWs at 8 K. This suggests that the carriers in the QWs are basically redistributed across the SMWR structure, thus contributing to increase the emission intensity. Finally, we found that having a top and bottom AlInGaP barrier with Al content around 80 % (tensile strained) increased the overall RT intensity by a factor ~2 (see figure S6a-b).

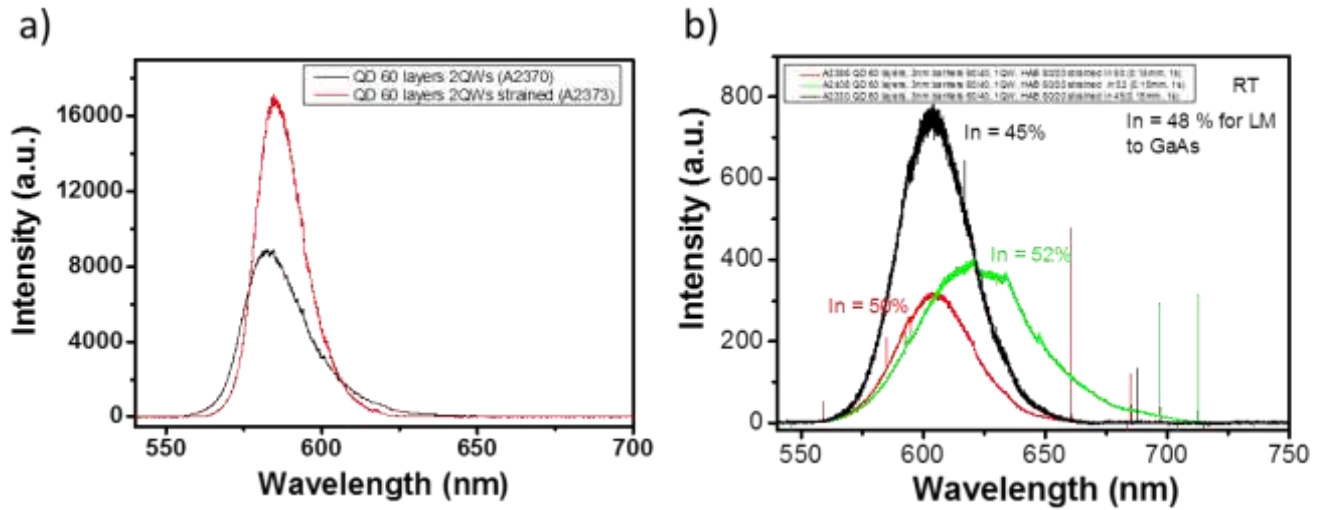

**Figure S5**: (a) Photoluminescence spectra of two identical QW-SMWR structures with and without top and bottom strained $(Al_{0.8}Ga_{0.2})_{0.52}In_{0.48}P$ barriers. (b) Comparison between nominally identical QW-SMWR structures with tensile and compressive strained $(Al_{0.8}Ga_{0.2})_{0.52}In_{0.48}P$ barriers.

## 8. Possible wire formation mechanism

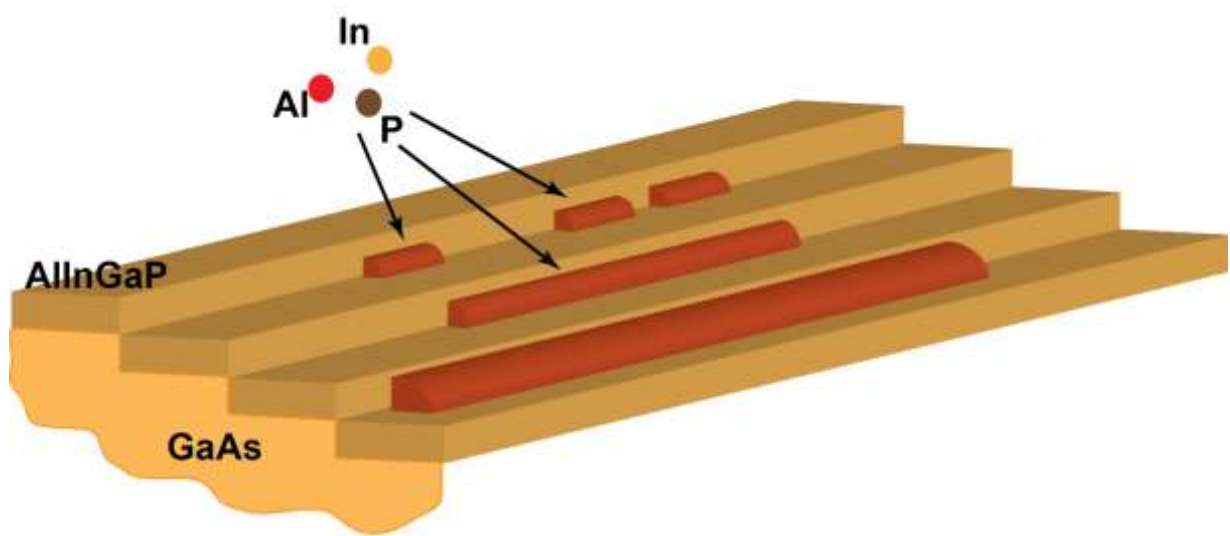

**Figure S6**: Schematic representation of a possible growth mechanism of AlInP wires. The step-bunched AlInGaP surface offers a natural template that drives the elongation of the wires.